\documentclass[a4paper,11pt]{article}
\usepackage{amsmath,amsfonts,amssymb}
\usepackage{pos}
\usepackage{slashed}
\usepackage{caption}
\usepackage{subcaption}

\newcommand{\gf}{\gamma_5}
\newcommand{\mui}{\mu_I}
\newcommand{\avg}[1]{\left\langle {#1} \right\rangle}
\newcommand{\Tr}{\mathrm{Tr}}
\newcommand{\mpcac}{m_\mathrm{pcac}}
\newcommand{\mud}{m_\mathrm{ud}}

\title{Pion condensation at non-zero isospin chemical potential with Wilson fermions}

\author*[a]{Rocco Francesco Basta}
\author[b]{Bastian B. Brandt}
\author[a]{Francesca Cuteri}
\author[b,c]{Gergely Endrődi}
\author[a]{Owe Philipsen}

\affiliation[a]{Institut für Theoretische Physik, Goethe-Universität Frankfurt,\\
  Max-von-Laue-Straße 1, 60438 Frankfurt am Main, Germany}

\affiliation[b]{Institute for Theoretical Physics, University of Bielefeld, D-33615 Bielefeld, Germany}

\affiliation[c]{Institute of Physics and Astronomy,
ELTE E\"otv\"os Lor\'and University,\\
P\'azm\'any P.\ s\'et\'any 1/A, H-1117 Budapest, Hungary}

\emailAdd{basta@itp.uni-frankfurt.de}

\abstract{In contrast to the case of non-zero baryon chemical potential, the isospin chemical potential does not introduce a sign problem and can be simulated on the lattice. When the isospin chemical potential is large enough, a phase transition to a Bose-Einstein condensate of pions takes place. Currently available results in the literature on the phase diagram and the equation of state in this setup employ staggered fermions. We present preliminary results on the onset of the pion condensation phase in simulations with Wilson fermions.}

\FullConference{The 41st International Symposium on Lattice Field Theory (LATTICE2024)\\
 28 July - 3 August 2024\\
Liverpool, UK\\}

\begin{document}
\maketitle

\section{Introduction}

\begin{figure}
    \centering
    \includegraphics[width=0.5\linewidth]{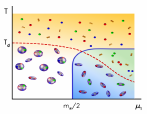}
    \caption{Schematic sketch of the conjectured phase diagram of QCD at $\mu_B = 0, \,\mui \neq 0$ in the $(T,\mui)$ plane, taken from \cite{Brandt:2017oyy}. The color scheme is the following: hadron phase (white), separated by the Quark-Gluon Plasma (yellow) by the thermal crossover (red dashed line), and by the pion condensation phase (blue). A further BCS phase (green) is predicted by perturbation theory.}
    \label{fig:phase-diagram}
\end{figure}

The grand canonical partition function of QCD depends on the temperature $T$ and the quark chemical potentials $\mu_q$. When we consider only up and down quarks, it is customary to rewrite them in terms of
\begin{equation}
    \mu_B \equiv  \frac{3(\mu_\mathrm{u} + \mu_\mathrm{d})}{2} \quad\text{and}\quad \mu_I \equiv \frac{\mu_\mathrm{u} - \mu_\mathrm{d}}{2} \,,
\end{equation}
where $\mu_B$ is the \emph{baryon} chemical potential and $\mu_I$ is the \emph{isospin} chemical potential. Examples of systems with non-zero $\mu_I$ are nuclear matter inside neutron stars, and neutron-rich nuclei in heavy ion collisions.

The study of QCD at non-zero $\mu_B$ on the lattice is complicated by the \emph{sign problem}, which only allows Monte Carlo simulations at $\mu_B = 0$. Switching on $\mui$ does not cause the same issues, allowing the use of importance sampling to study the properties of strong interacting matter at non-zero isospin densities. This system has been previously tackled with a variety of approaches, from chiral perturbation theory \cite{Son:2000xc} to lattice simulations of the grand canonical partition function with staggered fermions \cite{Kogut:2002tm,Kogut:2002zg,Kogut:2004zg,Endrodi:2014lja,Brandt:2017oyy,Cuteri:2021hiq,Brandt:2022_EoS,Brandt:2023kev}, and canonical methods on the lattice \cite{deForcrand:2007uz,Detmold:2012wc,Abbott:2023coj,Abbott:2024vhj}. A sketch of the phase diagram is shown in Fig.~\ref{fig:phase-diagram}. The $SU(2)_V$ isospin symmetry is broken explicitly by $\mui$. At low temperatures, the residual $U_{\tau_3}(1)$ symmetry breaks spontaneously for $\mui > \mu_{I,c}$, where $\mu_{I,c}(T=0) = m_\pi/2$, and a Bose-Einstein condensate (BEC) of pions appears (\emph{pion condensation phase}, in blue). 

Simulations in the grand canonical formalism have been previously performed using staggered quarks. The study of spectroscopy \cite{Rindlisbacher2014, Nonaka2013} and of the phase diagram \cite{Nakamura:2003gj,Nonaka2014} with Wilson fermions has been attempted before. Here, we present preliminary results on the onset of the pion condensation phase using Wilson fermions at larger than physical quark masses.

\section{Simulation setup}
\subsection{Action and fermion determinant}
We simulate QCD with two quark flavours at non-zero $\mu_I$ with the Wilson plaquette action and unimproved Wilson fermions. The action is given by
\begin{equation}
    S = S_W + \overline{\psi} \mathcal{M}_\mathrm{ud} \psi \quad\text{with}\quad \psi = \begin{pmatrix}
                    u \\
                    d
                \end{pmatrix}
    \quad\text{and}\quad
    \mathcal{M}_\mathrm{ud} = \begin{pmatrix}
                    D(\mui) & \lambda \gf \\
                    - \lambda \gf & D(-\mui)
                \end{pmatrix} \,,
\end{equation} 
where $D(\mui) \equiv \slashed{D}(\mui) + m_0$ is the massive Dirac operator with non-zero chemical potential, and $\lambda$ is the \emph{pion source term}. In order to observe spontaneous symmetry breaking on a finite lattice, it is necessary to perform simulations at $\lambda > 0$ and then extrapolate $\lambda \to 0$ in order to recover the correct physics. Moreover, spontaneous symmetry breaking causes near-zero modes of $D(\mui)$ to appear, which makes the use of $\lambda$ as an infrared regulator necessary for the stability of the simulations. 

The determinant can be written as
\begin{equation}
    \det \mathcal{M}_\mathrm{ud} = \det [D(\mui)^\dagger D(\mui) + \lambda^2]
\end{equation}
which is real and, when $\lambda > 0$, positive. As anticipated before, there is no sign problem in this setup.

We use the SMD update algorithm \cite{Horowitz:1985kd,Horowitz:1986dt,Horowitz:1991rr,Jansen:1995gz,Francis:2019muy} and use a rational approximation \cite{Clark:2003na} for the square root appearing in the pseudofermion heatbath step in combination with Hasenbusch preconditioning~\cite{Hasenbusch:2001ne, Hasenbusch:2002ai, Luscher:2012av}, writing the fermion determinant as
\begin{equation}
    \det \mathcal{M}_\mathrm{ud} = {W \det R^2[D(\mui)^\dagger D(\mui) + \lambda^2_n] \prod_{k=0}^{n-1} \frac{\det R^2[D(\mui)^\dagger D(\mui) + \lambda^2_{k}]}{\det R^2[D(\mui)^\dagger D(\mui) + \lambda^2_{k+1}]}} \quad\quad R[x] \simeq \sqrt{x}
\end{equation}
where $W = \det \{[D(\mui)^\dagger D(\mui) + \lambda_0^2]/R^2[D(\mui)^\dagger D(\mui) + \lambda_0^2]\}$ is a reweighting factor measured stochastically (in a similar way to \cite{Bruno:2014jqa}), and $\lambda \equiv \lambda_0 < \lambda_1 < \ldots < \lambda_{n}$ is a monotonically increasing sequence of pion source terms. Our simulation code is based on \texttt{openQCD-2.0} \cite{openQCD}.

\subsection{Observables}
In order to look at the boundary of the pion condensation phase, we study the unrenormalized pion condensate
\begin{equation}
    \pi^\pm \equiv \frac{\partial \log Z}{\partial \lambda} = \frac{2T}{V} \Tr \frac{\lambda}{D^\dagger D + \lambda^2}
\end{equation}
as the order parameter of the BEC phase transition. Since simulations have to be performed at $\lambda > 0$, results have to be extrapolated to $\lambda \to 0$ after performing the thermodynamical limit $V \to \infty$. To help with the extrapolation, an improved formulation was introduced in \cite{Brandt:2017oyy} in terms of the singular values $\xi_n$ of the Dirac operator $D(\mui)$:
\begin{equation}
    D(\mui)^\dagger D(\mui) \psi_n = \xi^2_n \psi_n.
\end{equation}
With this basis, one obtains a Banks-Casher relation for the pion condensate by writing
\begin{align}
    \pi^\pm &= \frac{2T}{V} \Tr \frac{\lambda}{D^\dagger D + \lambda^2} = \frac{2T}{V} \sum_n \frac{\lambda}{\xi_n^2 + \lambda^2} \nonumber \\
            &\xrightarrow[V \to \infty]{} 2\lambda \avg{\int d\xi \rho(\xi) \frac{1}{\xi^2 + \lambda^2}} 
             \xrightarrow[\lambda \to 0]{} \pi \avg{\rho(0)}.
\end{align}
This relation can be used to define an improved pion condensate operator, which has been shown to be less dependent on $\lambda$ and $V$, allowing a linear extrapolation to $\lambda \to 0$ on simulations with staggered fermions.

The singular value density $\avg{\rho(0)}$ can be measured from the \emph{averaged spectral density} $N(\xi)/\xi$, as
\begin{equation}
    \lim_{\xi \to 0} \frac{\pi}{N_\tau N_\sigma^3} \frac{N(\xi)}{\xi} = \pi \avg{\rho(0)}  
\end{equation}
where $N(\xi)$ is the integrated spectral density
\begin{equation}
        N(\xi) \equiv \int_0^{\xi} \rho(\xi') d\xi'.
\end{equation}
For each ensemble $(\mui, \lambda)$, the measurement of $\avg{\rho(0)}$ is carried out as follows: The first $100$ singular values of $D(\mui)$ are measured on each gauge configuration using the Krylov-Schur algorithm as implemented in \texttt{SLEPc} \cite{Hernandez:2005:SSF}. Then, one computes $N(\xi)/\xi$ by building a histogram with bin size $\Delta \xi$. The low end of $N(\xi)/\xi$ is extrapolated to $\xi \to 0$ using different polynomial models and fit regions, weighing each fit with the resulting $\chi^2$ and with the number of degrees of freedom of the fit. One can then build another histogram with the results of these fits in order to obtain $\avg{\rho(0)}$ and its statistical uncertainty. Since the result may vary depending on the resolution $\Delta \xi$ used when building the first histogram, the procedure is repeated with different bin sizes $\Delta \xi$ in order to obtain a more robust estimate of $\avg{\rho(0)}$ and its error. 
    
\section{Results}
In this section, we consider a $8 \times N_\sigma^3$ lattice with bare parameters $\beta = 4.9228$, $\kappa = 0.1815$, corresponding to $a = 0.311(3)$ fm, $m_{\pi} = 560(6)$ MeV and $T \simeq 79$ MeV, taken from \cite{Philipsen:2016hkv}. 

\subsection{$\lambda$ and volume dependence of the improved pion condensate}
\begin{figure}
    \centering
    \includegraphics[width=0.7\linewidth]{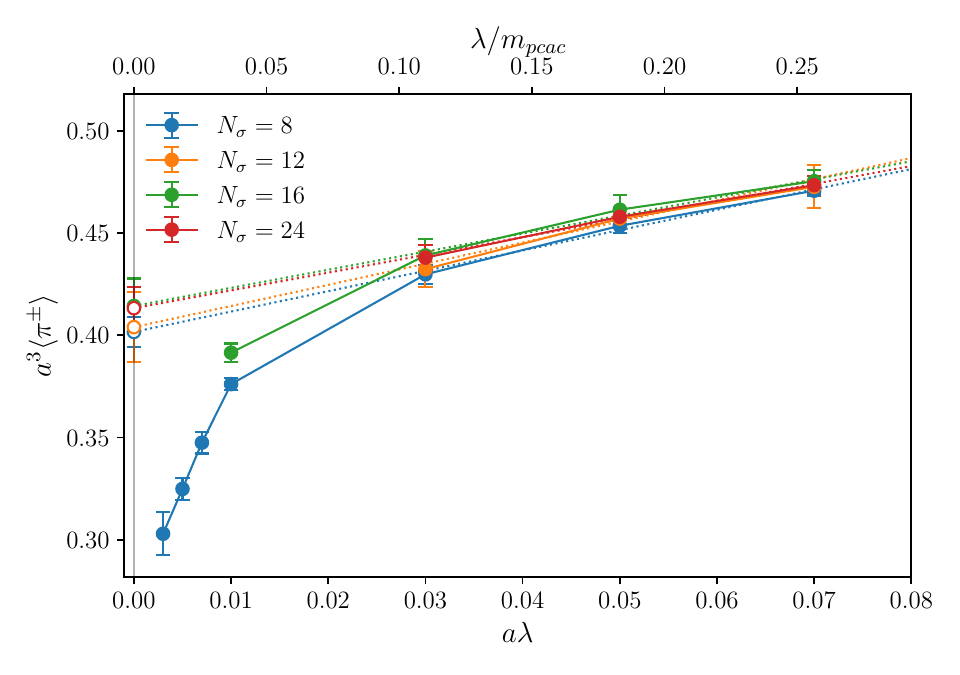}
    \caption{Improved pion condensate at $\mui \simeq 0.7 m_\pi$ on a $8\times N_\sigma^3$ lattice as a function of $\lambda$ and $N_\sigma$.}
    \label{fig:lambda-volume-dependence}
\end{figure}

Simulations must be performed at non-zero $\lambda$. But which values should we choose? To answer this question, we fix the isospin chemical potential to $\mu_I = 0.7 m_\pi$, inside the pion condensation phase, and we study the dependence of the pion condensate on $\lambda$ and on the spatial volume ($N_\sigma = 8, 12, 16, 24)$. 

In the staggered case, one compares $\lambda$ with the bare quark mass $\mud$, and the extrapolation was checked in \cite{Brandt:2017oyy} down to $\lambda/\mud \gtrsim 0.1$. In the Wilson case, the quark mass renormalizes additively due to the explicit breaking of chiral symmetry, and its computation requires the knowledge of the critical value of $\kappa_c(\beta)$ at which the pion mass vanishes. Due to this difficulty, we instead consider the PCAC mass $\mpcac$ \cite{Luscher:1996sc} defined at zero $T$, $\mui$ and $\lambda$, which is related to the subtracted bare quark mass $\mud$ by a multiplicative renormalization constant \cite{Fritzsch:2010aw, Brandt:2016daq}. We measure $\mpcac$ on a $24^4$ lattice on $O(200)$ configurations using wall sources, and $a\mpcac = 0.27305(23)$.

In principle, the physical case is recovered for $\lambda \to 0$. However, $\lambda \neq 0$ causes the pions to have a non-zero mass. If $\lambda$ is too small, the pions are too light and their correlation lengths become much larger than the spatial size of the lattice, causing large finite-size effects in the measurements. 

The results are shown in Fig.~\ref{fig:lambda-volume-dependence}. When $\lambda/\mpcac \gtrsim 0.1$ one can perform a linear extrapolation to $\lambda \to 0$, and the measurements are fairly insensitive to $N_\sigma$. For smaller values of $\lambda$, this is no longer the case: the behavior is not linear anymore, and measurements at different $N_\sigma$ disagree with each other, showing significant finite-size effects.

\subsection{The BEC phase boundary at $T \simeq 79$ MeV}

\begin{figure}[h]
    \centering
    \begin{subfigure}[b]{0.49\textwidth}
         \centering
         \includegraphics[width=\textwidth]{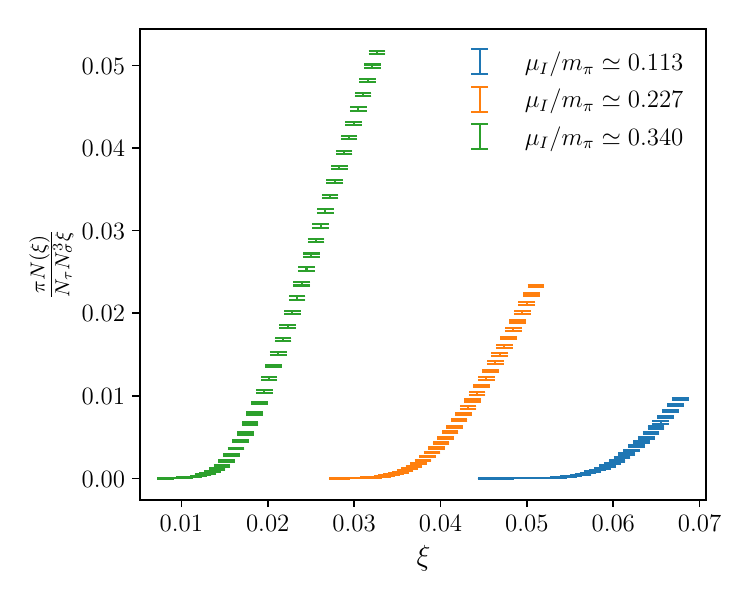}
         \caption{$\mu_I < m_\pi / 2$}
         \label{fig:int-spectral-density-below}
     \end{subfigure}
     \begin{subfigure}[b]{0.49\textwidth}
         \centering
         \includegraphics[width=\textwidth]{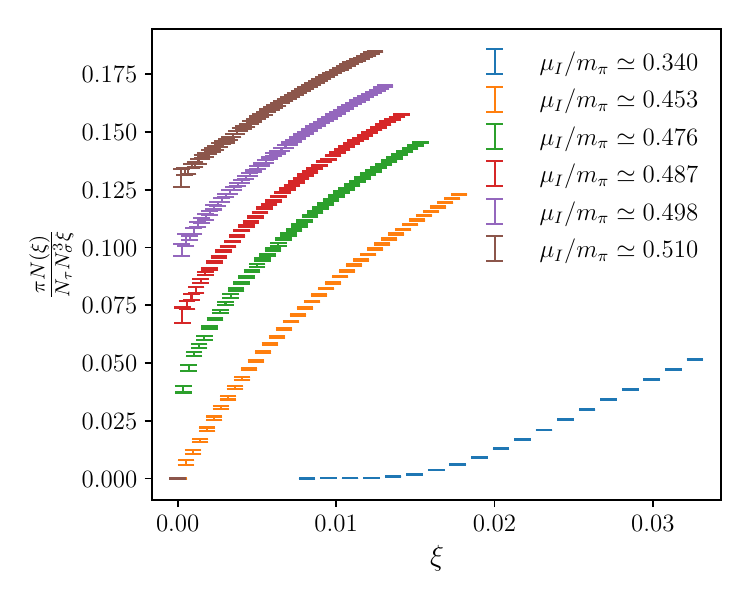}
         \caption{$\mu_I \sim m_\pi / 2$}
         \label{fig:int-spectral-density-transition}
     \end{subfigure}
    \caption{Averaged spectral density $\pi N(\xi)/(N_\tau N_\sigma^3\xi)$ at fixed $a\lambda = 0.05$ ($\lambda/\mpcac \simeq 0.18$) and varying $\mui$, before (left) and across (right) the BEC transition.}
    \label{fig:int-spectral-density}
\end{figure}

Let us now look at the BEC transition. We fix $N_\sigma = 24$ and, for now, $a\lambda = 0.05$. We show in Fig.~\ref{fig:int-spectral-density} the averaged spectral density
\begin{equation}
    h(\xi) \equiv \frac{\pi}{N_\tau N_\sigma^3} \frac{N(\xi)}{\xi} 
\end{equation}
which in the $\xi \to 0$ limit gives us the improved pion condensate
\begin{equation}
    \lim_{\xi \to 0} h(\xi) = \pi \avg{\rho(0)} \equiv \avg{\pi^\pm} 
\end{equation}
When $\mu_I < m_\pi / 2$, the histogram admits no extrapolation to a positive value, indicating that the density of singular values $\avg{\rho(0)}$ is zero. As we increase $\mu_I$, the histogram moves towards smaller and smaller eigenvalues, until we cross the transition and the low modes of $D(\mu_I)^\dagger D(\mu_I)$ start to accumulate for $\mu_I/m_\pi \gtrsim 0.47$, causing the $\xi \to 0$ extrapolation to be non-negative. We note that no $\lambda \to 0$ extrapolation has been performed yet.

\begin{figure}
    \centering
    \includegraphics[width=0.7\linewidth]{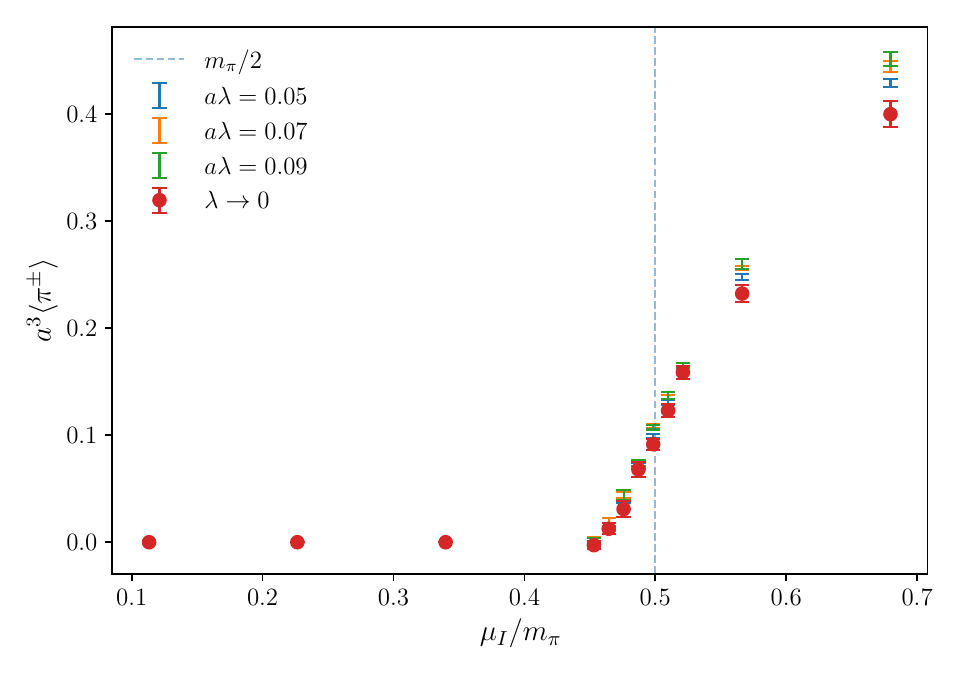}
    \caption{Improved pion condensate as a function of $\mui/m_\pi$, including the $\lambda \to 0$ extrapolation.}
    \label{fig:pion-condensate}
\end{figure}

\begin{figure}
    \centering
    \begin{subfigure}[b]{0.49\textwidth}
         \centering
         \includegraphics[width=\textwidth]{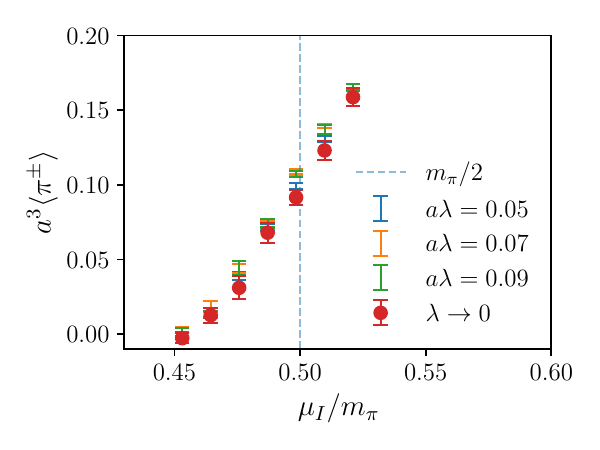}
         \caption{Our data}
         \label{fig:pion-condensate-zoom}
     \end{subfigure}
     \begin{subfigure}[b]{0.49\textwidth}
         \centering
         \includegraphics[width=\textwidth]{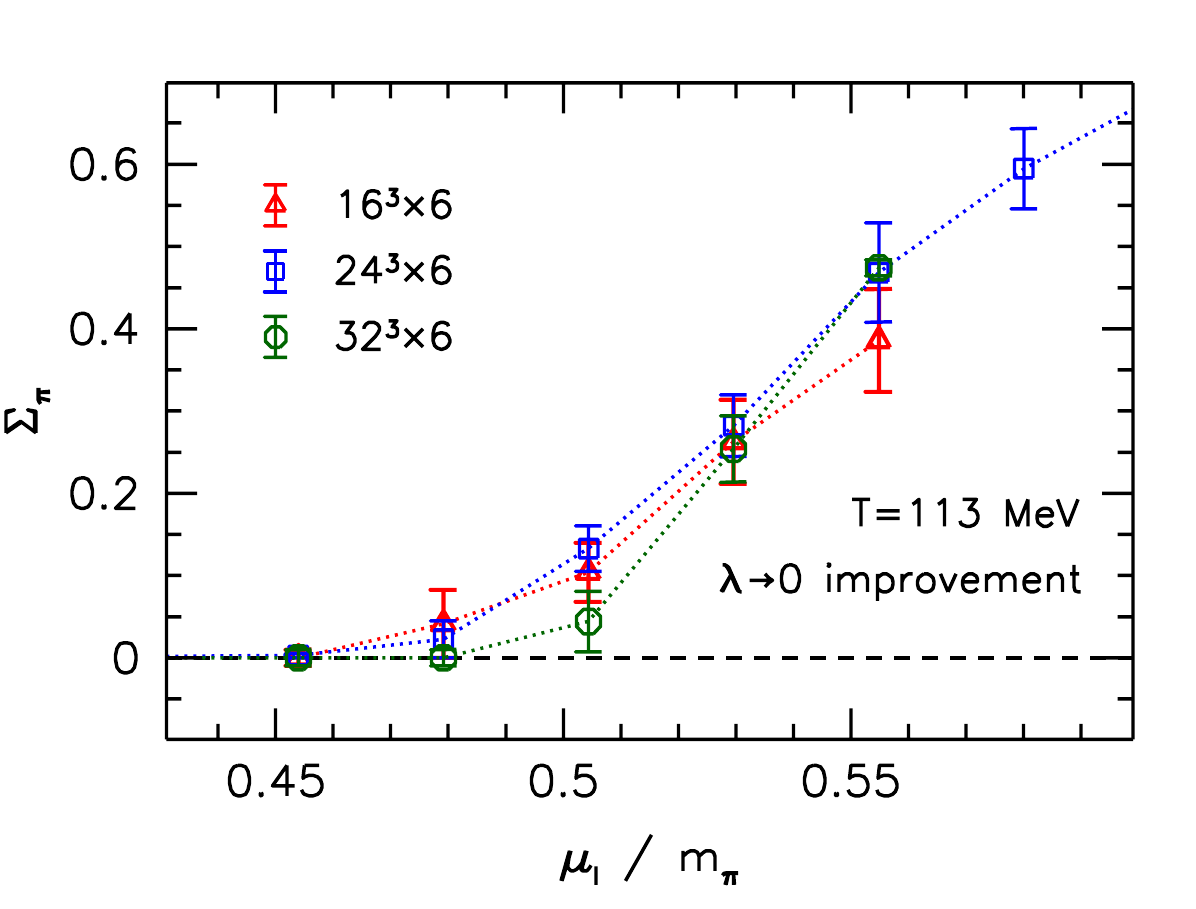}
         \caption{Staggered fermions \cite{Brandt:2017oyy}}
         \label{fig:pion-condensate-staggered}
     \end{subfigure}
    \caption{(a): Same as Fig.~\ref{fig:pion-condensate}, but only showing a narrow region around the transition. (b): Similar plot on a different setup, using staggered fermions (from \cite{Brandt:2017oyy}).}
    \label{fig:pion-condensate-zoom-comparison}
\end{figure}

We consider then $a\lambda = 0.05, 0.07, 0.09$ and we show the pion condensate at various $\mu_I$, with the corresponding $\lambda \to 0$ extrapolations, in Fig.~\ref{fig:pion-condensate}. The pion condensate becomes non-zero around the transition, which starts to happen slightly before the expected continuum prediction of $\mu_{I, c} = m_\pi /2$ which is expected to hold in the thermodynamical and continuum limit. We show a comparison with staggered results from \cite{Brandt:2017oyy} in Fig.~\ref{fig:pion-condensate-zoom-comparison}. Despite a multiplicative factor difference due to renormalization (which has not been carried out in the Wilson case), both plots show a similar behavior. 

\section{Conclusions and outlook}
We simulate two-flavour QCD with a non-zero isospin chemical potential on the lattice using Wilson fermions at $T \simeq 79$ MeV, $m_\pi = 560(6)$ MeV and $a = 0.311(3)$ fm. Using an improved definition \cite{Brandt:2017oyy} of the pion condensate as an order parameter, we first check its dependency on $\lambda$ and on the spatial volume, and then observe the transition to the pion condensation phase for $\mui \gtrsim m_\pi/2$. The onset of the transition is slightly shifted to the left of the expected value $\mu_{I, c} = m_\pi/2$, and this discrepancy is expected to be caused by finite size effects and lattice artifacts. We plan to extend this study further into the phase diagram, and to study the equation of state in this setup. 

\acknowledgments

RFB and OP acknowledge the support by the State of Hesse within the Research Cluster ELEMENTS (Project ID 500/10.006). 
RFB, BB, GE and OP acknowledge support by the Deutsche Forschungsgemeinschaft (DFG, German Research Foundation) through the CRC-TR 211 'Strong-interaction matter under extreme conditions'– project number 315477589 – TRR 211.
GE acknowledges funding by the Hungarian National Research, Development and Innovation Office (Research Grant Hungary 150241) and the European Research Council (Consolidator Grant 101125637 CoStaMM).
RFB acknowledges the support of the Helmholtz Graduate School for Hadron and Ion Research. 
The authors acknowledge the use of the Goethe-NHR cluster of the Center for Scientific Computing at the Goethe University Frankfurt and thank the computing staff for their support. We also acknowledge the use of the simulation code \texttt{openQCD} \cite{openQCD} and thank its authors.

\bibliographystyle{JHEPmod}
\bibliography{references.bib}

\end{document}